\documentclass{ws-procs9x6}

\begin{document}

\title{Magnetic Monopoles in Hot QCD }

\author{C. P. Korthals Altes}

\address{Centre Physique Th\'eorique au CNRS,\\
Luminy, 13288 Marseille Cedex, France\\
$^*$E-mail: ab\_altes@cpt.univ-mrs.fr\\
}


\begin{abstract}
 In this talk we review how a dilute gas of  magnetic monopoles in the adjoint
 describes  the spatial k-Wilson loops. We formulate an effective theory from
  $S_{MQCD}$ by integrating out dof's down to scales in between the magnetic
 screening mass and the string tension and relate the 3d pressure and the
 string tension.  Lattice data are consistent with the  gas being dilute for
all temperatures. 
  
\end{abstract}


\bodymatter

\section{Motivation and outline}\label{intro}

The QCD plasma phase consists of  colour electric quasi-particles,
gluons, whose interactions get screened by the Debye effect. Their flux is not
confined, like in a glueball. 

The plasma  is a nice theoretical laboratory to understand effects of
quasi-particles on flux-loops. Flux-loops are either of the colour-electric
variety (spatial 't Hooft loop) or of the magnetic variety (spatial Wilson
loop).   

Why do we mention ``magnetic''? In QED plasmas there is no reason to suspect
magnetic quasi-particles because long range static magnetic fields are
possible, in contrast to the screening of long range electric fields. But in
QCD there is indeed magnetic screening, as seen by lattice simulations.
So there must be magnetic activity in the QCD plasma. That is corroborated 
by the behaviour of the  spatial  Wilson loop: through Stokes law it measures
the flux of magnetic quasi-particles (MQP's or  ``monopoles'')> Its thermal
average behaves like an area law, as expected from a gas of free screened 
magnetic charges. In contrast to electric screening the magnetic screening is
non-perturbative. The same is valid for the respective loops.

These magnetic charges are supposed to be in a representation of a global
magnetic SU(N) group. They are non-perturbative objects. In order to produce a
pressure proportional to $N^2-1$ (valid to all orders in perturbation theory)
the monopoles be better in the adjoint representation. And this is enough
to compute the tension for the Wilson loop in the totally antisymmetric
representation built from  k quark representations. Its highest weight $Y_k$
controls the Stokes form for the loop. The adjoint
representation has a multiplicity $2k(N-k)$ of non-zero charges $\pm 2\pi/g$
in $Y_k|_{adj}$.  Every magnetic charge contributes the same
irrespective of the sign, and independently. Hence the tension is proportional
to this multiplicity
and this is up to 1 or 2 percent reproduced by the lattice data.
In the last section we  set up an effective theory for the monopole field. 

\section{Dimensional reduction, and the magnetic sector}\label{dim}

Dimensional reduction is valid at temperatures well above $T_c$. We will
be brief and only mention the essentials. Notation is g for the coupling, and 
N for the number of colours.Potentials are NxN matrices.
 
For the purpose of computing  magnetic loops it reduction is  particularly suitable.
This is because the original 4d QCD action reduces at high T to 3d 
Magnetostatic QCD, the 3d Yang Mills theory:
\begin{eqnarray}
S_{QCD}&\rightarrow & S_{EQCD}\rightarrow S_{MQCD},\\
 S_{EQCD}&=&(\vec D A_0)^2+m_D^2A_0^2+\lambda_E A_0^4 +F_{ij}^2+ ...\\
S_{MQCD}&=&F_{ij}^2+ ...
\label{reduction}
\end{eqnarray}

Both of the reduction steps are perturbative. The first integrates out
the hard T modes, leaving only static modes. The resulting Electrostatic QCD
Lagrangian is three dimensional and contains the $A_0$ potential as a massive
excitation with the Debye mass $m_D^2=(N/3)g^2T^2$, and of the 3d Yang-Mills
  action, with the typical magnetic scale $g^2T$. The second step is valid
when the Debye mass is much larger than the  magnetic scale $g^2T$. Then we can
integrate perturbatively the $A_0$ potential, and what stays is the 3d
Yang-Mills action $S_{MQCD}$. This action has no perturbative components left
at the scale $g^2T$.

\subsection{Magnetic observables}

These are the screening of the force
between two Dirac Z(N) monopoles~\cite{deforaltes} and the spatial Wilson loop.
The first gives a temperature dependent screening mass $m_M(T)$:
\nopagebreak
\begin{equation}
V_M(r)=d{\exp{(-m_Mr)}\over r}.
\label{screening} 
\end{equation}
It can be shown~\cite{deforaltes}  that this mass equals the mass gap in a 4d
SU(N) gauge theory with one periodic space dimension of period $1/T$. This
relation is valid for all $T$, in particular $T=0$. At very large $T$ this
mass can be simulated in 3d $S_{MQCD}$, a considerable simplification. 

The same magnetic activity that causes screening can be monitored by the
spatial Wilson loop, that measures the average colour magnetic flux.
The loop is carrying a representation $R$ built from a number $k$ of quark
representations.  Then the tension $\sigma_k$ is defined by:
\begin{equation}
\langle W_R(C)\rangle=\int DA_0D\vec A\exp{(ig\oint\vec A_R.d\vec
  l)}\sim\exp(-\sigma_k A)
\label{tension} 
\end{equation}
\noindent A the area of the minimal surface subtended by C.

In three and four dimensions $\sigma_k$ only depends on the N-allity $k$
~\footnote{In three dimensions a proof exists~\cite{meyer06}. It is based on
  the confining ground states being due to global magnetic Z(N) being
  broken. The unique tension $\sigma_k$ is that of the domain wall separating 
two vacua differing by the phase $\exp{(ik{2\pi\over N})}$. In 4d there is a
consensus, based on circumstantial proof.}.
So we need to discuss only one representation of N-allity $k$, and we will
limit ourselves to the fully antisymmetric one, built from k boxes. Its
highest weight  $Y_k$, written as a traceless NxN matrix, equals:
\begin{equation}
Y_k={1\over N}\mbox{diag}(k,k,...,k,k-N,k-N,...,k-N) 
\label{hypercharge} 
\end{equation}
This highest weight generalizes hypercharge.
The average of the loop in this representation admits a Stokes law:
\begin{equation}
\langle W_k(C)\rangle=\int DA_0 D\vec A D\Omega\exp{\big(ig\int d\vec S.Tr\big(\vec
  B-{i\over g}\vec D(A)Y_k^{\Omega}\times\vec D(A)Y_k^{\Omega}\big)Y_k^{\Omega}\big)}.
\label{stokes} 
\end{equation}
The gauge transformation $\Omega$ acts on $Y_k$ as on a frozen Higgs field:
$Y_k^{\Omega}= \Omega Y_k \Omega^{\dagger}$. As such this integral will
 only converge when we admit a Higgs field that fluctuates around its frozen
 value~\cite{diakonov}, i.e. a Higgs phase with the unbroken components
 $SU(k)\times SU(N-k)\times U(1)$. It is known from simulations that these are
 the only stable broken phases~\cite{rajantie}.   In the gauge $\Omega=1$ the
commutator terms in $\vec B$ and in the second covariant derivative term do
cancel. This   property determines  the $Y_k$ uniquely.  For
the surviving Abelian expression Stokes theorem is true and a gauge invariant 
projection of the line integral results. For the relation of this line
integral to the original 
Wilson loop see ref.~\cite{diakonov}.

Both magnetic screening and tension are at high T dominated by the 3d
$S_{MQCD}$.

\subsection{Electric observables}
Electric observables are the familiar Debye screening mass $m_D$ between two heavy
quarks, and the mean flux as measured by the spatial 't Hooft loop.

The spatial 't Hooft loop, given by a macroscopic closed Z(N) magnetic vortex~\cite{thooft78}
can be written as a flux loop in the physical Hilbert space:
\begin{equation}
V_k(C)=\exp{(i{4\pi\over g}\int d\vec S.Tr\vec E Y_k)}.
\label{eloop} 
\end{equation}

The surface in the integral is subtended by $C$.

Note the similarity to the first term in eq.(\ref{stokes}). 

The average of the loop obeys an area law in the deconfined phase, due to the
free colour charges in the plasma:
\begin{equation}
\langle V_k\rangle=\exp{(-\rho_k A))}
\label{etension} 
\end{equation}

\section{Theoretical results compared to lattice data}

The electric loop can be computed in perturbation theory through
integration of the hard modes and Debye modes\cite{bhatta}. This has been done 
including order $O(g^3)$. Including effects of hard modes to order $g^4$ is
desirable as will become evident below. 
The analytic results give Casimir scaling $k(N-k)$ for the contribution of
the hard modes.

To get an  insight in the analytic results one
 considers a quasi-particle gas of Debye screened longitudinal gluons
with density $n$ per gluon species in the octet (or adjoint for general N).
One species should contribute to the tension proportional to its density n.
To get the dimension for the tension right one should scale n by $m_D$. 
 In the quasi-particle picture this scaling is simply due to the
multiplicity  $2k(N-k)$ of gluons in the adjoint with non-zero $Y_k$
charge: $\rho_k/m_D^2\sim k(N-k)n/m_D^3\sim 1/(g\sqrt N)^3>>1$, the weak
 coupling plasma condition.

Below are
shown the data for various groups, with Casimir scaling divided out. Remarkably
a universal curve results dow to near $T_c$! Doing the same for the
theoretical result gives the curve. Including the  hard corrections to $g^4$ for
 the coupling~\cite{york2004} brings the theoretical curve down. Hence the need
 for $g^4$ for the tension itself.

\begin{figure}[htb]
\centerline{\epsfig{file=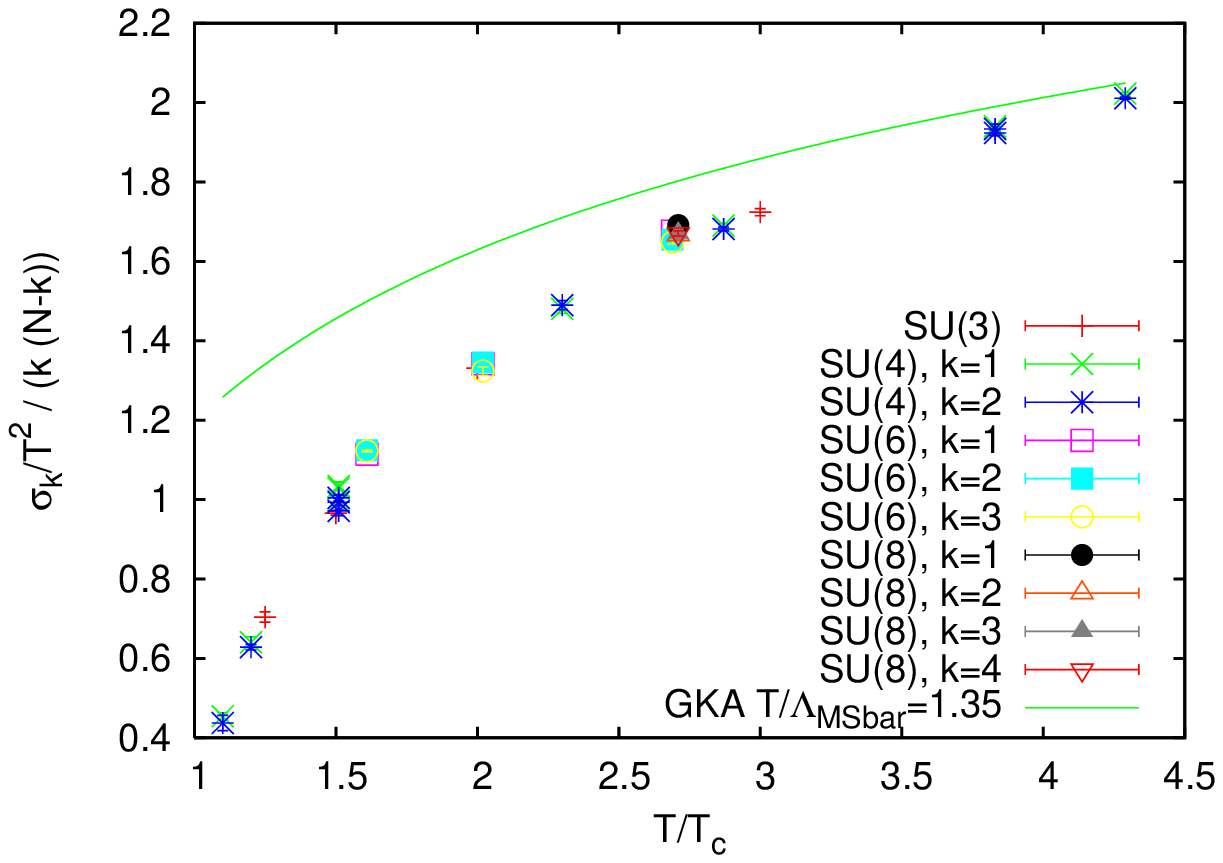,width=8.5cm}}
\centerline{The reduced e-tension for $SU(N_c)$, $N_c\le 8$, PdF  et al.,hep-lat/0510081}
\end{figure}

Turning to the magnetic loop, we are faced with its dominant behaviour being
due to the non-perturbative magnetic sector. For lack of analytic calculations
we proposed~\cite{giovanna2001} a magnetic quasi-particle model that should
explain qualitatively the pressure. As the pressure scales  like $N^2-1$ to
all orders in perturbation theory an adjoint multiplet of dilute monopoles is needed.
With an adjoint the same multiplcity argument as for the electric loop is
valid (eq.\ref{stokes}). So the k-tension of the Wilson loop has again Casimir
scaling:
\begin{equation}
\sigma_k\sim k(N-k) n_M/m_M.
\label{ktensionscales}
\end{equation}

The density of a given species is $n_M$, the screening of the monopoles is
$m_M$. We will take it as their size.  If that is the case then:
\begin{equation}  
{\sigma_1\over{m_M^2}}\sim (N-1){n_M\over {m_M^3}}.
\label{diluteness}
\end{equation}

In striking contrast to the weak coupling plasma condition in the electric case in the previous subsection, the left hand side of this relation is for all N a
{\it{ small}} number, $\sim 0.05$~\cite{teper98}. So for finite N the diluteness $\delta={n_M\over{m_M^3}}$ is small. For large N the l.h.s. is $O(1)$, hence $n_M=O(1/N)$. Below are shown the 3d data, which are at most off by two percent.
 
\begin{figure}[htb]
\centerline{\epsfig{file=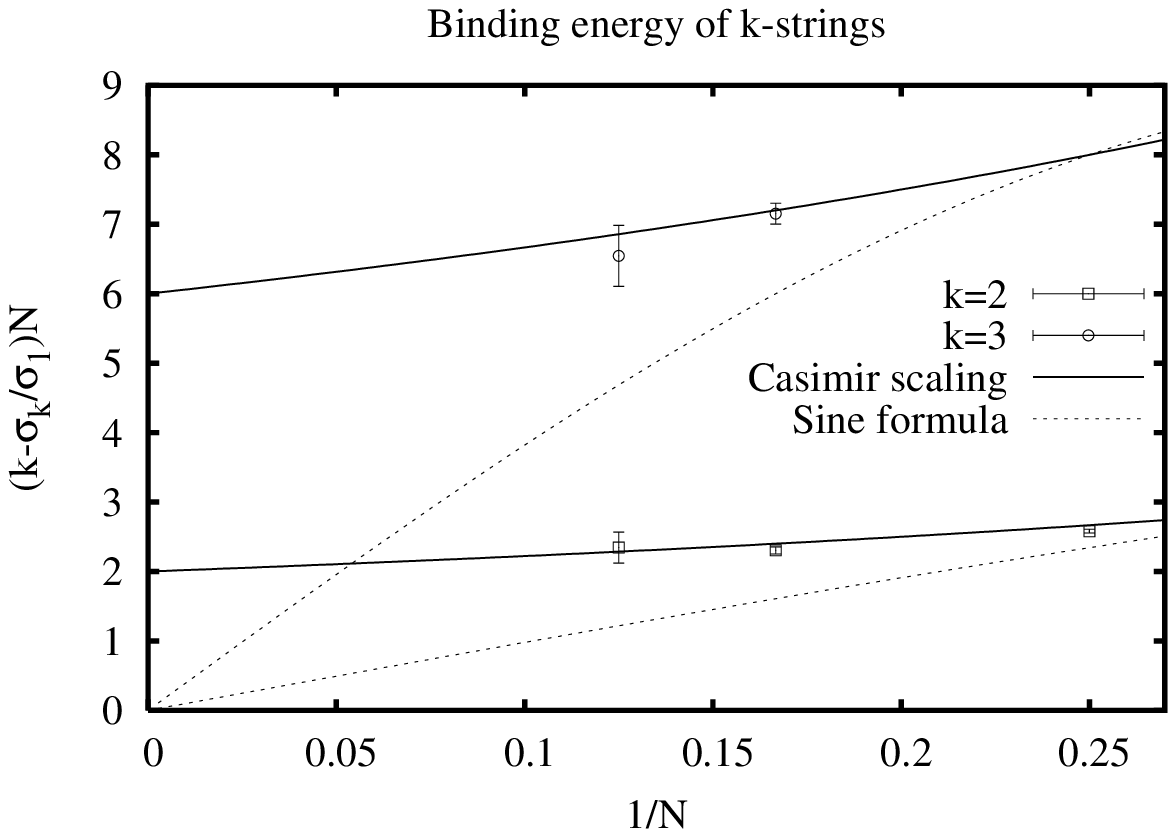,width=8.5cm}}
\centerline{m- flux-tension for $SU(N_c)$, $N_c\le 8$; Meyer, hep-lat/0412021) }
\end{figure}

\section{Effective action for the monopoles}
 
The raison d'\^etre of the monopole  model is that it gives the right group
factor for pressure and Wilson loops. At the same time the diluteness is
insured by the empirically small ratio $\sigma_1/m_M^2$. This warrants a more 
methodical approach. 

The question is whether one can define an effective action starting from
$S_{MQCD}$. The new action, $S_{MQP}$, follows by integrating out all scales
smaller or equal to the size of the monopole, $m_M$. The monopole field $M$
is a traceless hermitean field, transforming under a magnetic global SU(N)
group: $M\rightarrow UMU^{\dagger}$.


The new action should be invariant under this group. This invariance was not
there in the magnetostatic action. There are many examples of  a symmetry 
appearing in an effective action and being  broken by the higher order terms.

 With  $S_{MQP}$ we should be able
to compute the k-tensions and the pressure. When computing the pressure we
will use the $\overline{MS}$ scheme.
 The effective action reads in terms of M scaled by a mass $\sqrt{m_0}$:
\begin{equation}
S_{MQP}={g_3^6N^3(N^2-1)\over {(4\pi)^4}}\big(ylog(\bar\mu/2g_3N) +c\big) +
m_0 Tr[M(-\partial^2M + m_0^2)(1-\cos M)]+....
\label{MQP}
\end{equation}
\noindent where the potential term is that for a dilute Coulomb gas, in analogy
with Polyakov~\cite{thooft78}.  

The first term is the perturbative 4 loop contribution.
The coefficient y is known and equals
${43\over{12}}-{157\over{768}}\pi^2$~\cite{york2004}.

 $\bar\mu$ is a scale yet to be fixed by perturbative matching. The constant
  and  the mass term $m_0= O(g_3^2N)$ are unknown.  Doing the  integration
  over  $M$ gives then the pressure:
 \begin{equation} 
p_{\overline MS}={g_3^6N^3(N^2-1)\over {(4\pi)^4}}\big(({43\over{12}}-{157\over{768}}\pi^2)log(\bar\mu/2g_3N)+B_G+O(\epsilon)\big).
\label{pressure}
\end{equation}
\noindent where $B_G=c+ f(m_0/g_3^2N)$.

The short distance behaviour of the cosine potential should be consistent with the first  term capturing all of it.
A recent lattice determination~\cite{direnzo}  gives a result
consistent with zero for $B_G$. A straightforward computation of the tension $\sigma_1$ fixes $m_0$.

\section{Conclusions}
We argued that not only electric quasi-particles- the longitudinal gluons
but also magnetic quasi-particles (with spin zero as well) explain nicely the flux loop data. These data  do not admit dyonic quasi-particles, which would
strongly correlate the loops. 

What happens to the model when we go down in temperature?

The variation of the ratio $\sigma_1/m_M^2$ from very high T to T=0 is from about
$0.05$ to $0.09$. So the gas can still be considered dilute at T=0. If so, $T_c$ would
the transition where the Bose gas becomes superfluid, and the magnetic
symmetry gets spontaneously broken. Vortices in the superfluid become the
locus of electric flux strings. Above $T_c$ they are absent, so  the Polyakov loop develops a VEV above $T_c$.
 One should have experimental signals for  this superfluid phase.

I thank Mikko Laine and Harvey Meyer for discussions.

\nopagebreak

\end{document}